\documentclass[runningheads,fleqn]{cl2emult}
\usepackage{graphicx}
\usepackage{subeqnar}
\usepackage{multicol}
\usepackage{cropmark}
\usepackage{engi}
\usepackage{amsfonts}
\usepackage{amsmath}

\begin{document}

\title*{A non-linear hardening model based on two coupled internal hardening
variables: formulation and implementation }

\toctitle{A non-linear hardening model: formulation and
implementation}
%
%
\titlerunning{A non-linear hardening model }
%

\author{Nelly Point\inst{1 ,}\inst{2}
\and Silvano Erlicher\inst{1 ,}\inst{3}}

\authorrunning{Nelly Point and Silvano Erlicher}

\institute{Ecole Nationale des Ponts et Chauss\'{e}es,\\
         Laboratoire d'Analyse des Mat\'{e}riaux et Identification,\\
         6-8 avenue Blaise Pascal,\\
         Cit\'{e} Descartes, Champs-sur-Marne,\\
         F-77455 Marne la Vall\'{e}e Cedex 2, France
\and Conservatoire National des Arts et M\'{e}tiers,\\
         D\'{e}partement de Math\'{e}matiques,\\
         292 rue Saint Martin,\\
         F-75141 Paris Cedex 03, France\\
         E-mail: point@cnam.fr
 \and  Universit\`{a} di Trento, \\
        Dipartimento di Ingegneria Meccanica e Strutturale  \\
        Via Mesiano 77, 38050, Trento, Italy \\
        E-mail: silvano.erlicher@ing.unitn.it}
\maketitle

\begin{abstract}
An elasto-plasticity model with coupled hardening variables of
strain type is presented. In the theoretical framework of
generalized associativity, the formulation of this model is based
on the introduction of two hardening variables with a coupled
evolution. Even if the corresponding hardening rules are linear,
the stress-strain hardening evolution is non-linear. The numerical
implementation by a standard return mapping algorithm is discussed
and some numerical simulations of cyclic behaviour in the
univariate case are presented.

\end{abstract}

\section{Introduction}

Starting from the analysis of the dislocation phenomenon in metallic
materials, Zarka and Casier \cite{ZarkaCasier79} and Kabhou et al. \cite%
{Kabhou85} proposed an elasto-plasticity model ("four-parameter model")
where, in addition to the usual kinematic hardening internal variable, a
second strain like internal variable was introduced. It plays a role in a
modified definition of the von Mises criterion and its evolution, defined by
linear flow rules, is coupled with the one of the kinematic hardening
variable. The resulting elasto-plastic model depends only on four
parameters. Its non-linear hardening behavior was studied in \cite%
{Inglebert99} and a parameter identification method using essentially a
cyclic uniaxial test was presented in \cite{PointVial2000} . In this note,
the thermodynamic formulation of the classical elasto-plastic model with
linear kinematic and isotropic hardening is first recalled. Then, by using
the same theoretical framework, a generalization of the four-parameter model
is suggested, relying on the introduction of an additional isotropic
hardening variable. Finally, a return mapping implementation of the
generalized model is presented and some numerical simulations are briefly
discussed.

\section{Thermodynamic formulation of a plasticity model with linear
kinematic/isotropic hardening}

Under the assumption of isothermal infinitesimal transformations and of
isotropic material, the hydrostatic and the deviatoric responses can be
treated separately (see, among others, \cite{LemaitreChaboche90}). Hence,
the free energy density $\Psi $ can be split into its spherical part $\Psi
_{h}$ and its deviatoric part $\Psi _{d}$. To obtain \emph{linear} state
equations, $\Psi _{h}$ and $\Psi _{d}$ are assumed quadratic. Moreover,
experimental results for metals show that permanent strain is only due to
deviatoric slip. Hence, an elastic spherical behaviour is assumed, leading
to the following definition :%
\begin{equation}
\Psi _{h}=\frac{1}{2}\left( \lambda +\frac{2\mu }{3}\right) tr\left( \mathbf{%
\varepsilon }\right) ^{2}=\frac{1}{2}K\text{ }tr\left( \mathbf{\varepsilon }%
\right) ^{2}  \label{HelmSpher}
\end{equation}%
where $\mathbf{\varepsilon }$ is the (small) strain tensor, $\lambda $ and $%
\mu $ are the Lam\'{e} constants and $K$ is the bulk modulus. Under the same
assumptions, the deviatoric potential $\Psi _{d}$ must depend only on
deviatoric state variables. The plastic flow is associated to the plastic
strain $\mathbf{\varepsilon }^{p}$, while the kinematic/isotropic hardening
behaviour is introduced by the tensorial internal variable $\mathbf{\alpha }$
and by the scalar variable $p$ :%
\begin{equation}
\Psi _{d}=\Psi _{d}\left( \mathbf{\varepsilon }_{d}\mathbf{,\varepsilon }%
^{p},\mathbf{\alpha },p\right) =\frac{2\mu }{2}\left( \mathbf{\varepsilon }%
_{d}-\mathbf{\varepsilon }^{p}\right) :\left( \mathbf{\varepsilon }_{d}-%
\mathbf{\varepsilon }^{p}\right) +\frac{B}{2}\text{ }\mathbf{\alpha :\alpha +%
}\frac{H}{2}p^{2}  \label{HelmDeviat}
\end{equation}%
where $tr\left( \mathbf{\alpha }\right) =tr\left( \mathbf{\varepsilon }%
^{p}\right) =0$ and $B,\ H>0$ . The evolution of $p$ will be related
to the norm of $\mathbf{\varepsilon }^{p}$. \newline
The state equation concerning the deviatoric stress tensor is easily derived:%
\begin{equation}
\mathbf{\sigma }_{d}=\frac{\partial \Psi _{d}}{\partial \mathbf{\varepsilon }%
_{d}}=2\mu \left( \mathbf{\varepsilon }_{d}-\mathbf{\varepsilon }^{p}\right)
\label{StateEq}
\end{equation}%
and the thermodynamic forces associated to $\mathbf{\varepsilon }^{p},$ $%
\mathbf{\alpha }$ and $p$ are defined by :%
\begin{equation}
\left\{
\begin{array}{l}
\mathbf{\sigma }_{d}=-\frac{\partial \Psi _{d}}{\partial \mathbf{\varepsilon
}^{p}}=2\mu \left( \mathbf{\varepsilon }_{d}-\mathbf{\varepsilon }^{p}\right)
\\
\mathbf{X}=\frac{\partial \Psi _{d}}{\partial \mathbf{\alpha }}=B\text{ }%
\mathbf{\alpha } \\
R=\frac{\partial \Psi _{d}}{\partial p}=K\text{ }p%
\end{array}%
\right.  \label{ForceThermo}
\end{equation}%
One can notice that $tr\left(\mathbf{\varepsilon }\right)=0$. The
linearity of the hardening rules (\ref{ForceThermo})$_{2-3}$ follows
from the quadratic form assumed for the last two terms in (\ref{HelmDeviat}%
). The second principle of thermodynamics can be written as follows \cite%
{LemaitreChaboche90} :%
\begin{equation}
\mathbf{\sigma }_{d}:\mathbf{\dot{\varepsilon}}_{d}-\dot{\Psi}_{d}\geq 0
\label{2ndPrinciple1}
\end{equation}%
By using (\ref{ForceThermo}) in (\ref{2ndPrinciple1}), the
Clausius Duhem
inequality is obtained :%
\begin{equation}
\mathbf{\sigma }_{d}:\mathbf{\dot{\varepsilon}}^{p}-\mathbf{X}:\mathbf{\dot{%
\alpha}}-R\text{ }\dot{p}\geq 0  \label{2ndPrinciple2}
\end{equation}%
In order to fulfil this inequality, a classical assumption is to
impose that
$(\mathbf{\dot{\varepsilon}}^{p},\mathbf{\dot{\alpha}},$
$\dot{p})$ belongs to the subdifferential of a positive convex
function $\phi _{d}^{\ast }$, equal to zero in zero, called
pseudo-potential. In such a case, the evolution of the internal
variables is compatible with (\ref{2ndPrinciple2})
\cite{HalphenNguyen75}.

The von Mises criterion corresponds to a special choice for the
pseudo-potential $\phi _{d}^{\ast }\left( \mathbf{\sigma }_{d},\mathbf{X}%
,R\right) $ which is equal, in this case, to the indicator function $\mathbb{%
I}_{f\leq 0}$ of the elastic domain, or, to be more specific, of the set of $%
\left( \mathbf{\sigma }_{d},\mathbf{X},R\right) $ such that the
so-called
yielding function $f$ is non-positive :%
\begin{equation}
f=f\left( \mathbf{\sigma }_{d},\mathbf{X},R\right) =\left\Vert \mathbf{%
\sigma }_{d}-\mathbf{X}\right\Vert -\sqrt{\frac{2}{3}}\sigma
_{y}-R\leq 0 \label{yield funct}
\end{equation}%
where $\left\Vert \cdot \right\Vert $ is the standard
$L_{2}$-norm. To
impose that $(\mathbf{\dot{\varepsilon}}^{p},\mathbf{\dot{\alpha}},$ $\dot{p}%
)$ belongs to the subdifferential of $\mathbb{I}_{f\leq 0}$ is
equivalent to
write :%
\begin{equation}
\left\{
\begin{array}{l}
\mathbf{\dot{\varepsilon}}^{p}=\dot{\lambda}\frac{\partial
f}{\partial
\mathbf{\sigma }_{d}}=\dot{\lambda}\frac{\mathbf{\sigma }_{d}-\mathbf{X}}{%
\left\Vert \mathbf{\sigma }_{d}-\mathbf{X}\right\Vert } \\
\mathbf{\dot{\alpha}}=-\dot{\lambda}\frac{\partial f}{\partial \mathbf{X}}=%
\dot{\lambda}\frac{\mathbf{\sigma }_{d}-\mathbf{X}}{\left\Vert \mathbf{%
\sigma }_{d}-\mathbf{X}\right\Vert } \\
\dot{p}=-\dot{\lambda}\frac{\partial f}{\partial R}=\dot{\lambda}%
\end{array}%
\right.   \label{normality}
\end{equation}%
with the conditions $\dot{\lambda}\geq 0$ $,$ \ $f\leq 0$ \ and\ $\dot{%
\lambda}f=0$ . \newline Equations (\ref{normality}) are called
generalized associativity conditions or associative flow rules.
The relations (\ref{normality}) yield in this
case $\mathbf{\dot{\alpha}}\ =$ $\mathbf{\dot{\varepsilon}}^{p}$ and$\ \dot{%
\lambda}=\dot{p}=\left\Vert
\mathbf{\dot{\varepsilon}}^{p}\right\Vert $.
\newline
>From (\ref{normality})$_{2}$ and (\ref{ForceThermo})$_{2}$ one
obtains the Prager's linear kinematic hardening rule and a linear
isotropic hardening
rule :%
\begin{equation}
\mathbf{\dot{X}}=B\text{ }\mathbf{\dot{\varepsilon}}^{p}\text{ \ ,
\ \ \ \ \ }\dot{R}=H\text{ }\dot{p}  \label{linear hardening}
\end{equation}\newline%
The coefficient $\dot{\lambda}$ is strictly positive only if
$f=0.$ In this
case, its value can be derived from the so-called consistency condition $%
\dot{f}=0$, i.e.%
\begin{equation*}
\frac{\partial f}{\partial \mathbf{\sigma }_{d}}:\mathbf{\dot{\sigma}}_{d}%
\mathbf{+}\frac{\partial f}{\partial \mathbf{X}}:\mathbf{\dot{X}}+\frac{%
\partial f}{\partial R}\text{ }\dot{R}\text{ }\mathbf{=}\text{ }0
\end{equation*}%
The introduction into the previous equation of the state equation (\ref{StateEq}),
as well as the thermodynamic force definitions (\ref%
{ForceThermo})$_{2-3}$ and the normality (\ref{normality}), yield :%
\begin{equation}
\frac{\partial f}{\partial \mathbf{\sigma }_{d}}:\mathbf{\dot{\sigma}}_{d}%
\mathbf{-}\dot{\lambda}\text{ }B\frac{\partial f}{\partial \mathbf{X}}:\frac{%
\partial f}{\partial \mathbf{X}}-\dot{\lambda}\text{ }H\frac{\partial f}{%
\partial R}\frac{\partial f}{\partial R}\mathbf{=}0
\label{ConsistThermForces}
\end{equation}%
Moreover, in a strain driven approach, Eq.
(\ref{ConsistThermForces}) has to be rewritten still using the
state equation (\ref{StateEq}). As a result, by
collecting $\dot{\lambda}$, one obtains%
\begin{equation*}
\dot{\lambda}=\mathcal{H}\left( f\right) \frac{2\mu \left\langle \frac{%
\partial f}{\partial \mathbf{\sigma }_{d}}:\mathbf{\dot{\varepsilon}}%
_{d}\right\rangle }{2\mu \frac{\partial f}{\partial \mathbf{\sigma }_{d}}%
\text{:}\frac{\partial f}{\partial \mathbf{\sigma }_{d}}+B\frac{\partial f}{%
\partial \mathbf{X}}\text{:}\frac{\partial f}{\partial \mathbf{X}}+H\frac{%
\partial f}{\partial R}\frac{\partial f}{\partial R}}=\text{ }\frac{\mathcal{%
H}\left( f\right) }{1+\frac{B+K}{2\mu }}\frac{\left\langle \left( \mathbf{%
\sigma }_{d}-\mathbf{X}\right)
:\mathbf{\dot{\varepsilon}}_{d}\right\rangle }{\left\Vert
\mathbf{\sigma }_{d}-\mathbf{X}\right\Vert }\geq 0
\end{equation*}%
where $\mathcal{H}\left( f\right) $ is zero when $f<0$ and equal to $1$ for $%
f=0.$ The symbol $\langle .\rangle $ represents the MacCauley
brackets.
\section{A generalization of the four-parameter model}

The linear hardening model discussed previously is used here to
suggest a generalization of the 4-parameter model cited in the
introduction. The tensor $\mathbf{\alpha }$ into Eq.
(\ref{HelmDeviat}) is replaced by a couple of tensors
$(\mathbf{\alpha }_{1},\mathbf{\alpha }_{2})$. As a result, the
scalar constant $B$ becomes a $2\times 2$ symmetric positive
definite matrix, denoted by $\mathbf{B=[}b_{ij}\mathbf{]}$. For
sake of simplicity, only the thermodynamic potential $\Psi _{d}$
is considered here and it is
defined as:%
\begin{equation*}
\Psi _{d}\left( \mathbf{\varepsilon }_{d}\mathbf{,\varepsilon }^{p},\mathbf{%
\alpha }_{1},\mathbf{\alpha }_{2},p\right) =\frac{2\mu }{2}\left( \mathbf{%
\varepsilon }_{d}-\mathbf{\varepsilon }^{p}\right) \text{:}\left( \mathbf{%
\varepsilon }_{d}-\mathbf{\varepsilon }^{p}\right) +\frac{1}{2}\mathbf{%
\alpha }^{T}\ \mathbf{B\ \alpha }+\frac{H}{2}p^{2}
\end{equation*}%
where $\mu $ and $H$ have the same meaning as before and $\mathbf{\alpha }$
is the column vector defined as $\mathbf{\alpha }=[\mathbf{\alpha }_{1};%
\mathbf{\alpha }_{2}]$. The state equation becomes :%
\begin{equation}
\mathbf{\sigma }_{d}=\frac{\partial \Psi _{d}}{\partial \mathbf{\varepsilon }%
_{d}}=2\mu \left( \mathbf{\varepsilon }_{d}-\mathbf{\varepsilon }^{p}\right)
\label{StateEqNew}
\end{equation}%
and the thermodynamic forces have the following form :%
\begin{equation}
\left\{
\begin{array}{l}
\mathbf{\sigma }_{d}=-\frac{\partial \Psi _{d}}{\partial \mathbf{\varepsilon
}^{p}}=2\mu \left( \mathbf{\varepsilon }_{d}-\mathbf{\varepsilon }^{p}\right)
\\
\mathbf{X}_{1}=\frac{\partial \Psi _{d}}{\partial \mathbf{\alpha }_{1}}%
=b_{11}\text{ }\mathbf{\alpha }_{1}+b_{12}\text{ }\mathbf{\alpha }_{2}\text{
} \\
\mathbf{X}_{2}=\frac{\partial \Psi _{d}}{\partial \mathbf{\alpha }_{2}}%
=b_{21}\text{ }r\text{ }\mathbf{\alpha }_{1}+b_{22}\text{ }\mathbf{\alpha }%
_{2} \\
R=\frac{\partial \Psi _{d}}{\partial p}=H\text{ }p%
\end{array}%
\right. \qquad \text{or \ \ \ \ \ \ \ }\left\{
\begin{array}{l}
\mathbf{\sigma }_{d}=2\mu \left( \mathbf{\varepsilon }_{d}-\mathbf{%
\varepsilon }^{p}\right) \\
\mathbf{X}=\mathbf{B\ \alpha } \\
R=H\text{ }p%
\end{array}%
\right.  \label{ForceThermoNew}
\end{equation}%
where $\mathbf{X}=[\mathbf{X}_{1};%
\mathbf{X}_{2}]$. The Clausius -Duhem inequality becomes in this case :%
\begin{equation*}
\mathbf{\sigma }_{d}:\mathbf{\dot{\varepsilon}}^{p}-\mathbf{X}_{1}:\mathbf{%
\dot{\alpha}}_{1}-\mathbf{X}_{2}:\mathbf{\dot{\alpha}}_{2}-R\text{ }\dot{p}%
\geq 0\qquad \text{or \ \ \ \ \ \ \ }\mathbf{\sigma }_{d}:\mathbf{\dot{%
\varepsilon}}^{p}-\mathbf{X}^{T}\ \mathbf{\dot{\alpha}}-R\text{ }\dot{p}\geq
0
\end{equation*}%
Moreover, the loading function $f$ is defined as follows :%
\begin{equation}
f=f\left( \mathbf{\sigma }_{d},\mathbf{X}_{1},\mathbf{X}_{2},R\right) =\sqrt{%
\left\Vert \mathbf{\sigma }_{d}-\mathbf{X}_{1}\right\Vert ^{2}+\rho
^{2}\left\Vert \mathbf{X}_{2}\right\Vert ^{2}}-\sqrt{\frac{2}{3}}\sigma
_{y}-R\leq 0  \label{loadFunctNewModel}
\end{equation}%
with $\rho $ a positive scalar. One can remark that for $\rho =0$
and $H=0$ the standard von Mises criterion is derived while for
$\rho =1$ and $H=0$ the 4-parameter model is retrieved. The flow
rules are defined by a normality
condition :%
\begin{equation*}
\left( \mathbf{\dot{\varepsilon}}^{p},-\mathbf{\dot{\alpha}}_{1},-\mathbf{%
\dot{\alpha}}_{2},-\dot{p}\right) \in \partial \phi _{d}^{\ast }=\partial \mathbb{I}%
_{f\leq 0}
\end{equation*}%
Therefore, the proposed model belongs to the framework of generalized
associative plasticity \cite{HalphenNguyen75} . The loading function (\ref%
{loadFunctNewModel}) can be rewritten as \newline
$f=g\left( \mathbf{Y}_{1},\mathbf{Y}_{2}\right) -\sqrt{\frac{2}{3}}\sigma
_{y}-R\leq 0$ with $\mathbf{Y}_{1}=\mathbf{\sigma }_{d}-\mathbf{X}_{1}$ and $%
\mathbf{Y}_{2}=-\mathbf{X}_{2}$ . Hence%
\begin{equation}
\left\{
\begin{array}{l}
\mathbf{\dot{\varepsilon}}^{p}=\dot{\lambda}\frac{\partial f}{\partial
\mathbf{\sigma }_{d}}=\dot{\lambda}\frac{\mathbf{\sigma }_{d}-\mathbf{X}_{1}%
}{\sqrt{\left\Vert \mathbf{\sigma }_{d}-\mathbf{X}_{1}\right\Vert ^{2}+\rho
^{2}\left\Vert \mathbf{X}_{2}\right\Vert ^{2}}} \\
\mathbf{\dot{\alpha}}_{1}=-\dot{\lambda}\frac{\partial f}{\partial \mathbf{X}%
_{1}}=\dot{\lambda}\frac{\mathbf{\sigma }_{d}-\mathbf{X}_{1}}{\sqrt{%
\left\Vert \mathbf{\sigma }_{d}-\mathbf{X}_{1}\right\Vert ^{2}+\rho
^{2}\left\Vert \mathbf{X}_{2}\right\Vert ^{2}}} \\
\mathbf{\dot{\alpha}}_{2}=-\dot{\lambda}\frac{\partial f}{\partial \mathbf{X}%
_{2}}=-\dot{\lambda}\frac{\rho ^{2}\text{ }\mathbf{X}_{2}}{\sqrt{\left\Vert
\mathbf{\sigma }_{d}-\mathbf{X}_{1}\right\Vert ^{2}+\rho ^{2}\left\Vert
\mathbf{X}_{2}\right\Vert ^{2}}} \\
\dot{p}=-\dot{\lambda}\frac{\partial f}{\partial R}=\dot{\lambda}%
\end{array}%
\right. \text{ \ \ or \ \ }\left\{
\begin{array}{l}
\mathbf{\dot{\varepsilon}}^{p}=\dot{\lambda}\frac{\partial f}{\partial
\mathbf{\sigma }_{d}} \\
\mathbf{\dot{\alpha}}=\dot{\lambda}\nabla g \\
\dot{p}=\dot{\lambda}%
\end{array}%
\right.  \label{plasticFlow}
\end{equation}%
It can be seen from (\ref{plasticFlow}) that $\mathbf{\dot{\alpha}}_{1}=%
\mathbf{\dot{\varepsilon}}^{p}$ and $\dot{p}=\dot{\lambda}=\left\Vert
\mathbf{\dot{\alpha}}\right\Vert =\sqrt{\left\Vert \mathbf{\dot{\alpha}}%
_{1}\right\Vert ^{2}+\left\Vert \mathbf{\dot{\alpha}}_{2}\right\Vert ^{2}}.$
Moreover, from (\ref{plasticFlow})$_{1-2}$ and (\ref{ForceThermoNew})$_{2-4}$
one obtains the kinematic and isotropic hardening rules :%
\begin{equation*}
\mathbf{\dot{X}}_{1}=b_{11}\mathbf{\dot{\varepsilon}}^{p}+b_{12}\text{ }%
\mathbf{\dot{\alpha}}_{2}\text{, \ \ \ \ \ }\mathbf{\dot{X}}_{2}=b_{21}%
\mathbf{\dot{\varepsilon}}^{p}+b_{22}\text{ }\mathbf{\dot{\alpha}}_{2}\text{%
, \ \ \ \ \ }\dot{R}=H\text{ }\dot{p}
\end{equation*}%
In \cite{Inglebert99} it was proved that $\mathbf{B}$ can be written as :%
\begin{equation*}
\mathbf{B=}\left[
\begin{array}{cc}
\left( A_{\infty }+r^{2}b\right) & -rb \\
-rb & b%
\end{array}%
\right]
\end{equation*}%
where the scalars $A_{\infty }$ and \ $b$ are strictly positive and have the
dimension of stresses. For $r=0,$ there is no coupling and $\mathbf{B}$ is
diagonal, so that the dimensionless scalar $r$ can be seen as a coupling
factor in the evolutions of $\mathbf{X}_{1}$ and $\mathbf{X}_{2}$ :%
\begin{equation*}
\mathbf{\dot{X}}_{1}=A_{\infty \ }\mathbf{\dot{\varepsilon}}^{p}+rb\text{ }(r%
\mathbf{\dot{\varepsilon}}^{p}-\mathbf{\dot{\alpha}}_{2})\text{, \ \ }%
\mathbf{\dot{X}}_{2}=b(r\mathbf{\dot{\varepsilon}}^{p}-\text{ }\mathbf{\dot{%
\alpha}}_{2})\text{, \ }\Longrightarrow \text{\ \ }\mathbf{\dot{X}}_{1}+r%
\mathbf{\dot{X}}_{2}=A_{\infty \ }\mathbf{\dot{\varepsilon}}^{p}
\end{equation*}%
In the first two flow rules a recalling term appears, as in the non-linear
kinematic hardening model of Frederich and Armstrong \cite{FredArmstr66}. As
before, the plastic multiplier can be explicitly computed by the consistency
condition :%
\begin{equation*}
\dot{\lambda}=\mathcal{H}\left( f\right) \frac{\left\langle \frac{\partial f%
}{\partial \mathbf{\sigma }_{d}}:\mathbf{\dot{\varepsilon}}_{d}\right\rangle
}{1+\frac{\nabla g.\mathbf{B}.\nabla g+H}{2\mu }}\geq 0.
\end{equation*}

\section{Implementation and some numerical results}

In this section, a numerical implementation of the model is proposed. A
standard return mapping algorithm is considered (see \cite{SimoHughes86}).
The formulation is explicitly described in the univariate case, but the
tensorial generalization is straightforward. Let $\Delta t_{n}$ be the
amplitude of the time step defined by $t_{n}$ and $t_{n+1}$ and let $\mathbf{%
\tilde{\alpha}}_{n}=\left[ \alpha _{1,n},\alpha _{2,n},p_{n}\right] ^{T}$ \
and $\mathbf{\tilde{X}}_{n}=\left[ X_{1,n},X_{2,n},R_{n}\right] ^{T}$ be the
vectors collecting the internal variables and the corresponding
thermodynamic forces. Moreover, let%
\begin{equation*}
\begin{array}{l}
\mathbf{D}=\left[
\begin{array}{cc}
\mathbf{B} & \mathbf{0} \\
\mathbf{0} & H%
\end{array}%
\right]%
\end{array}%
\end{equation*}%
be the global hardening modulus matrix. In a strain driven approach, knowing
the value of all the variables at the time $t_{n}$ and the \emph{strain
increment \ }$\Delta \varepsilon _{n}$ occurring during the time step $%
t_{n}\rightarrow t_{n+1}$, the numerical scheme computes the variables value
at $t_{n+1}$:%
\begin{equation*}
\left( \varepsilon _{n},\varepsilon _{n}^{p},\mathbf{\tilde{\alpha}}%
_{n},\sigma _{n},\mathbf{\tilde{X}}_{n},f_{n}\right) +\Delta \varepsilon
_{n}\Longrightarrow \left( \varepsilon _{n+1},\varepsilon _{n+1}^{p},\mathbf{%
\tilde{\alpha}}_{n+1},\sigma _{n+1},\mathbf{\tilde{X}}_{n+1},f_{n+1}\right)
\end{equation*}%
The flow equations (\ref{plasticFlow}) define a first order differential
system, which can be solved by the implicit Euler method. Therefore, the
discrete form of the model evolution rules is (the notation $\partial _{%
\mathbf{w}}f$ is equivalent to $\partial f/\partial \mathbf{w}$) :

$f_{n+1}:=\sqrt{\left( \sigma _{n+1}\mathbf{-}X_{1,n+1}\right) ^{2}+\rho
^{2}\left( X_{2,n+1}\right) ^{2}}-\left( \sigma _{y}+R_{n+1}\right) \leq 0$

\begin{description}
\item $\sigma _{n+1}=E\left( \varepsilon _{n+1}-\varepsilon
_{n+1}^{p}\right) $ \ ; \ \ \ \ \ $\mathbf{\tilde{X}}_{n+1}=\mathbf{D\tilde{%
\alpha}}_{n+1}$ \ \ \ with $E=\mu \frac{3\lambda +2\mu }{\lambda +\mu }$

\item $\varepsilon _{n+1}^{p}-\varepsilon _{n}^{p}=\Delta \gamma _{n+1}$ $%
\partial _{\sigma }f_{n+1}$\ ; \ \ \ \ \ \ \ \ \ \ $\mathbf{\tilde{\alpha}}%
_{n+1}-\mathbf{\tilde{\alpha}}_{n}=-\Delta \gamma _{n+1}$ $\partial _{%
\mathbf{\tilde{X}}}f_{n+1}$

\item $\Delta \gamma _{n+1}\geq 0,$ \ \ \ \ $f_{n+1}\leq 0,$ \ \ \ \ $\Delta
\gamma _{n+1}$ $f_{n+1}=0.$
\end{description}

\begin{figure}[hb]
\includegraphics[width=11cm,height=8cm]{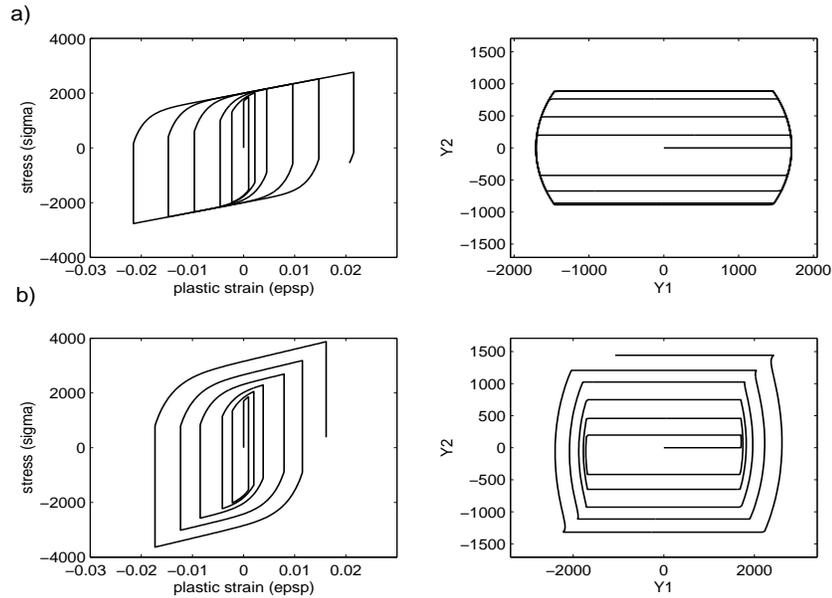}
\caption[]{Hysteresis loops for an imposed history with increasing
strain amplitude. a) $r=0.608,$ $H=0$ MPa b) $r=0.608,$ $H=6500$
MPa.}
 \label{fig1}
\end{figure}

\begin{figure}[h]
\includegraphics[width=11cm,height=8cm]{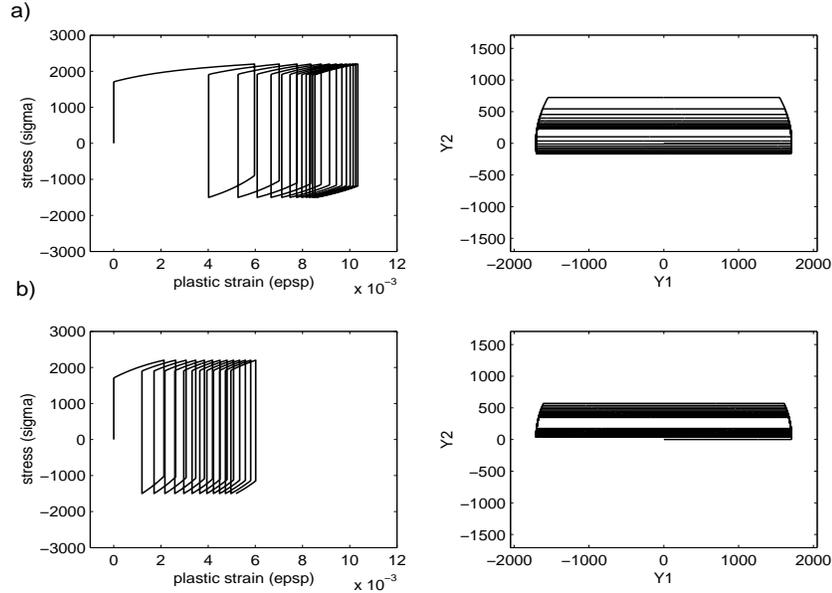}
\caption[]{Hysteresis loops for an imposed history with constant
stress amplitude and non-zero mean stress. a) $r=0.608,$ $H=0$ MPa$;$ b) $%
r=0.9,$ $H=0$ MPa.}
 \label{fig2}
\end{figure}

An elastic predictor-plastic corrector algorithm is used to take into
account the Kuhn-Tucker conditions \cite{SimoHughes86} (cf the last row). At
every time step, in the first predictor phase it holds $f_{n}<0$ , an
elastic behaviour is assumed and a trial value of $f_{n+1}$ , i.e. $%
f_{n+1}^{(0)}$, is computed. If $f_{n+1}^{(0)}\leq 0$ , then an elastic
behaviour occurs, $\Delta \gamma _{n+1}$ has to be zero and no corrector
phase is required. On the other hand, if $f_{n+1}^{(0)}>0$ , then plastic
strains occur, the elastic prediction has to be corrected and $\Delta \gamma
_{n+1}>0$ has to be computed. This is done by a suitable \emph{return
mapping algorithm}, described below :

\begin{description}
\item[i)] \textbf{Initialization}

\item $k=0;$ $\varepsilon _{n+1}^{p\left( 0\right) }=\varepsilon _{n}^{p},%
\mathbf{\tilde{\alpha}}_{n+1}^{\left( 0\right) }=\mathbf{\tilde{\alpha}}%
_{n},\gamma _{n+1}^{\left( 0\right) }=0$

\item[ii)] \textbf{Check yield condition and evaluate residuals}

\item $%
\begin{array}{l}
\sigma _{n+1}^{\left( k\right) }:=E\left( \varepsilon _{n+1}-\varepsilon
_{n+1}^{p\left( k\right) }\right) \text{ \ ; \ }\mathbf{\tilde{X}}%
_{n+1}^{\left( k\right) }:=\mathbf{D}\text{ }\mathbf{\tilde{\alpha}}%
_{n+1}^{\left( k\right) }\text{ \ ; \ }f_{n+1}^{\left( k\right) }:=f\left(
\sigma _{n+1}^{\left( k\right) },\mathbf{\tilde{X}}_{n+1}^{\left( k\right)
}\right) \\
\mathbf{R}_{n+1}^{\left( k\right) }:=\left[
\begin{array}{c}
-\varepsilon _{n+1}^{p\left( k\right) }+\varepsilon _{n}^{p} \\
\mathbf{\tilde{\alpha}}_{n+1}^{\left( k\right) }-\mathbf{\tilde{\alpha}}_{n}%
\end{array}%
\right] +\gamma _{n+1}^{\left( k\right) }\left[
\begin{array}{c}
\partial _{\sigma }f \\
\partial _{\mathbf{\tilde{X}}}f%
\end{array}%
\right] _{n+1}^{\left( k\right) } \\
\text{if: }f_{n+1}^{\left( k\right) }<tol_{1}\text{ \ \ \ \& \ \ \ }%
\left\Vert \mathbf{R}_{n+1}^{\left( k\right) }\right\Vert <tol_{2}\text{ \ \
\ then: \ EXIT}%
\end{array}%
$

\item[iii)] \textbf{Elastic moduli and consistent tangent moduli}

\item $%
\begin{array}{l}
C_{n+1}^{\left( k\right) }=E\text{ \ \ \ \ \ \ \ \ \ }\mathbf{D}%
_{n+1}^{\left( k\right) }=\mathbf{D} \\
\left( \mathbf{A}_{n+1}^{\left( k\right) }\right) ^{-1}=\left[
\begin{array}{cc}
\left( C_{n+1}^{-1}+\gamma _{n+1}\partial _{\sigma \sigma }^{2}f_{n+1}\right)
& \gamma _{n+1}\partial _{\sigma \mathbf{\tilde{X}}}^{2}f_{n+1} \\
\gamma _{n+1}\partial _{\mathbf{\tilde{X}}\sigma }^{2}f_{n+1} & \left(
\mathbf{D}_{n+1}^{-1}+\gamma _{n+1}\partial _{\mathbf{\tilde{X}\tilde{X}}%
}^{2}f_{n+1}\right)%
\end{array}%
\right] ^{\left( k\right) }%
\end{array}%
$

\item[iv)] \textbf{Increment of the consistency parameter}

\item $\Delta \gamma _{n+1}^{\left( k\right) }=\frac{f_{n+1}^{\left(
k\right) }-\left[
\begin{array}{cc}
\partial _{\mathbf{\sigma }}f_{n+1}^{\left( k\right) } & \partial _{\mathbf{%
\tilde{X}}}f_{n+1}^{\left( k\right) }%
\end{array}%
\right] ^{T}\mathbf{A}_{n+1}^{\left( k\right) }\mathbf{R}_{n+1}^{\left(
k\right) }}{\left[
\begin{array}{cc}
\partial _{\sigma }f_{n+1}^{\left( k\right) } & \partial _{\mathbf{\tilde{X}}%
}f_{n+1}^{\left( k\right) }%
\end{array}%
\right] ^{T}\mathbf{A}_{n+1}^{\left( k\right) }\left[
\begin{array}{cc}
\partial _{\sigma }f_{n+1}^{\left( k\right) } & \partial _{\mathbf{\tilde{X}}%
}f_{n+1}^{\left( k\right) }%
\end{array}%
\right] ^{T}}$

\item[v)] \textbf{Increments of plastic strain and internal variables}

\item $\left[
\begin{array}{c}
\Delta \varepsilon _{n+1}^{p\left( k\right) } \\
\Delta \mathbf{\tilde{\alpha}}_{n+1}^{\left( k\right) }%
\end{array}%
\right] =\left[
\begin{array}{cc}
C_{n+1}^{-1} & 0 \\
\mathbf{0} & -\mathbf{D}_{n+1}^{-1}%
\end{array}%
\right] ^{\left( k\right) }\mathbf{A}_{n+1}^{\left( k\right) }\left( \mathbf{%
R}_{n+1}^{\left( k\right) }+\Delta \gamma _{n+1}^{\left( k\right) }\left[
\begin{array}{c}
\partial _{\sigma }f_{n+1}^{\left( k\right) } \\
\partial _{\mathbf{\tilde{X}}}f_{n+1}^{\left( k\right) }%
\end{array}%
\right] \right) $

\item[vi)] \textbf{Update state variables and consistency parameter}

\item $\varepsilon _{n+1}^{p\left( k+1\right) }=\varepsilon _{n+1}^{p\left(
k\right) }+\Delta \varepsilon _{n+1}^{p\left( k\right) }$ \ ; \ \ $\mathbf{%
\tilde{\alpha}}_{n+1}^{\left( k+1\right) }=\mathbf{\tilde{\alpha}}%
_{n+1}^{\left( k\right) }+\Delta \mathbf{\tilde{\alpha}}_{n+1}^{\left(
k\right) }$ \ $\gamma _{n+1}^{\left( k+1\right) }=\gamma _{n+1}^{\left(
k\right) }+\Delta \gamma _{n+1}^{\left( k\right) }\bigskip $
\end{description}
This procedure to determine $\Delta \gamma _{n+1}$ requires the computation,
at each iteration, of the gradient and the Hessian matrix of $f$ . Other
algorithmic approaches by-pass the need of the Hessian of $f$ , but they are
not considered here.
\newpage
This implementation is used to obtain hysteresis loops in some
particular cases. The values of the four parameters $E,$ $\sigma
_{y},$ $A_{\infty }$ and $b$ are the same as those used in
\cite{PointVial2000} and correspond to the identified values of an
Inconel alloy ($E=205580$ Mpa$,$ $\sigma _{y}=1708,9$ Mpa$,$
$A_{\infty }=35500$ Mpa, $b=380700$ Mpa). The value of the new
parameter $\rho $ is $\rho =1$ and the values of $r$ and $H$ are
indicated in the caption of each figure.
 /newline
 Fig. \ref{fig1}
llustrates the hysteresis loops obtained with an increasing
amplitude strain history. The effect of the newly introduced
isotropic hardening term is highlighted. Fig. \ref{fig2} refers to
a stress input history, with constant amplitude and non-zero mean.
The plastic strain accumulation (ratchetting) and the shakedown
phenomenon are modelled by changing only one parameter. The
hysteresis loops are qualitatively similar to the ones of the
non-linear kinematic hardening model of Armstrong and Frederick
\cite{FredArmstr66}.

\section{Conclusions}

A model with coupled hardening variables of strain type has been
presented. It permits to take into account isotropic hardening and
to have an elastic unloading path of varying length depending on
the history of the loading. The simplicity of this model, which
depends only on six parameters, seems to be very attractive for
structural modelling applications with ratchetting effects. To
this aim, the proposed return mapping algorithm is a useful
numerical tool, which allows numerical simulations to be performed
in an effective way.

\end{document}